\documentclass[12pt,preprint]{aastex}
\usepackage{graphics}

\shorttitle{The line emission region in III Zw 2}

\shortauthors{Popovi\'c et al.}

\begin{document}

\title{The Line Emission Region in III Zw 2: Kinematics and Variability} 

\author{ L. \v C. Popovi\'c$^{1,2,3}$,  E. G. Mediavilla$^{4}$, E. 
Bon$^{1,3}$, 
N. Stani\'c$^{1,3}$ and  A. Kubi\v cela$^{1,3}$}

\affil{$^{1}$Astronomical  Observatory,  Volgina  7, 11160  Belgrade  74,
Serbia}

\affil{$^{2}$Astrophysikalisches Institut Potsdam, An der Sternwarte 16,
14482 
Potsdam, Germany (Alexander von Humboldt Fellow)}

\affil{$^{3}$Isaac Newton Institute of Chile,
Yugoslavia Branch (Casilla 8-9, Correo 9, Santiago, Chile)}

\affil{$^{4}$Instituto de  Astrof\'isica de Canarias C/ V\'ia L\`actea, 
s/n 
E38200 - La Laguna (Tenerife), Spain} 
\email{lpopovic@aob.bg.ac.yu}

\begin{abstract}
We  have  studied  the  Ly$\alpha$,  H$\beta$,  H$\alpha$ 
and   Mg II$\lambda$2798 line  profiles of the Seyfert 1 galaxy III Zw
2. The shapes of 
these
broad  emission lines show evidence of a multicomponent origin and also
features  which may be identified as the peaks due to a rotating disk. 
We have
proposed a  two-component Broad Line Region (BLR) model consisting of an
inner  Keplerian relativistic disk and  an outer structure surrounding 
the
disk. The results of the fitting of the  four Broad Emission Lines 
(BELs)
here considered, are highly consistent in both the inner and outer  component
parameters. Adopting a mass of  $\sim 2 \cdot 10^{8} \rm M_\odot$ for the
central object we found that the outer radius of the disk is approximately
equal for the four considered lines ($\sim 0.01$ pc). However, the inner
radius of the disk is not the same: 0.0018 pc for Ly$\alpha$, 0.0027 pc
for Mg
II, and 0.0038 pc for the Balmer lines. This as well as the relatively
broad
component present in the blue wings of the narrow [OIII] lines indicate
stratification in the emission-line region. Using long-term H$\beta$ 
observations (1972-1990, 1998) we found a flux variation of the BEL 
with respect to the [OIII] lines. 
 \end{abstract}


\section{Introduction} 

The  active  galaxy III  Zw 2  (Mrk  1501)  appears to  be  essentially 
stellar-like, with faint wisps  extending toward the northwest  (Arp 1968, 
Zwicky  1971). III Zw 2 presents the classic broad emission line characteristic 
of a type 1 Seyfert galaxy or a quasar (Arp 1968, Sargent 1970, Khachikian
\& 
Weedman 1974, Osterbrock 1977). The emission lines of III Zw 2 have been 
studied in several  papers   (Osterbrock 1977,   Kaastra  \& Korte 1988,
Corbin 
\& Borson 1996). Crenshaw et al. (1999)  noted the presence of  intrinsic 
absorption lines  in the ultraviolet spectrum of III Zw 2 obtained with 
the Hubble Space Telescope (HST). 

A disk model for the broad emission-line region of III Zw 2 has been 
proposed in  several papers (Kaastra \& 
Korte 1988, Corbin \& Boroson 1996, Shimura \& Takahara 1995,  Rokaki \&  
Boisson 1999).  The rotating 
accretion disk model (van Groningen 1983,  Kaastra \&
Korte 1988, Perez et al. 
1988, Chen et al. 1989, Chen \& Halpern 1989,  Halpern 1990, Eracleous \&
Halpern
1994, Pariev \& Bromley 
1998, Rokaki \& Boisson 1999, Shapovalova et al. 2001, Popovi\'c et al. 
2002, Kollatschny \& Bischoff 2002) has been  very
often discussed in order to 
explain the observed broad optical emission-line profiles in AGN.  This
model 
fits well  the widely 
accepted AGN paradigm in that the 'central engine'  consists of a 
massive black 
hole fueled by an  accretion disk. However, the fraction of AGN with 
double-peaked lines (which may indicate the disk emission in a line) is
small, 
and the observational evidence to support  the existence of a disk is not
statistically 
significant.

 Observations in a wide band of
wavelengths 
(X-ray, UV,  optical) also indicate that an accretion  disk could be 
present in III 
 Zw 2. In particular, Kaastra  and   Korte  (1988)  gave  some  
parameters  for  
the  central engine. They  hold that the accretion  disk has a  large 
 thickness in the central   part, that  it probably  extends to 0.2  pc 
and that 
 the central black  hole  has a mass of about  5$\cdot$10$^7$M$_\odot$. 
From the 
study of the emission lines,  Kaastra \&  Korte inferred dimensions of  $\sim 
7\cdot  10^{18}$ m and $\sim 10^{16}$ m for the NLR and the BLR, respectively. 
Shimura and Takahara (1995) reproduced the specific UV and soft X-ray 
luminosities in III Zw 2 from a disk emission spectrum.  More recently, 
Rokaki 
\& Boisson (1999) show that the UV continuum and the H$\beta$ line emission are 
compatible with an accretion disk model. 

The aim  of this paper is to analyze the shapes of the Hydrogen and  Mg 
II$\lambda$2798 emission lines of III Zw 2 in order to identify features 
which may be associated with the emission from a rotating disk and try to find  
evidence that suggests that 
disk emission can contribute to line emission. We  propose a 
structure for 
the BLR which can describe the shape of these lines. We  also use a set of 
observations of 
the H$\beta$  line,  taken over  a   long period of time, to discuss the
H$\beta$ 
line shape variability.

\section{Observations and data reduction} 

We  use spectra of III Zw 2 taken  from three sources: (i) 38 
spectra including the H$\beta$ line (wavelength interval 4750-5900 \AA) 
observed at the Crimean Astrophysical Observatory (CrAO) by K.K. Chuvaev
with the 
2.6 m Shain telescope during the period 1972-1990 (HJD 2441361  till 2448153),  
(ii) spectra taken with the HST in 1992 which include
the  
Ly$\alpha$  and MgII$\lambda 2798$ lines, and (iii) 3 spectra taken in 1998 
with the Isaac Newton Telescope (INT) at La Palma Observatory including
the
H$\alpha$  
and H$\beta$ lines. 

The spectra including the Ly$\alpha$ and  MgII$\lambda 2798$ lines were
also 
observed on January 18, 1992 with the HST Faint Object Spectrograph 
(FOS). Two
 different gratings were used to cover the wavelength range around the
Ly$\alpha$
and Mg II$\lambda 2798$ lines: a)   G130H in the spectral range from 1087.23\
\AA\ to 1605.52\ \AA\ with spectral resolution of 0.98\ \AA; and b) G270H
in
the spectral range from 2221.10\ \AA\ to 3300.10\ \AA\ with spectral
resolution
of 2.05\ \AA.  Three spectra of Ly$\alpha$ and one of Mg II$\lambda 2798$
were
taken. The spectra have been reduced by the HST team, and are in the format
intensity vs. wavelength. We obtained an averaged spectrum of Ly$\alpha$
from
the three observed.

At the CrAO, 38 spectra including the H$\beta$ line (wavelength interval
4750-5900 \AA) were obtained with the 2.6 m Shain telescope during the period
1972-1990. The spectral resolution was $\sim 8$\ \AA. The spectrograph slit and
 seeing were in the $1.8^{\prime\prime}-2.0^{\prime\prime}$,   and
$2^{\prime\prime}-3^{\prime\prime}$ ranges, respectively. The  spectra  of
H$\beta$   were scanned with a two-coordinate CrAO microphotometer (as in the
case of Akn 120, see e.g. Stani\'c  et al. 2000, Popovi\'c et al., 2001).  The
reduction procedure includes   corrections for  the  film sensitivity, sky
background,  and instrumental   spectral  sensitivity.   The wavelength and
flux calibration were made using  the SPE data reduction package, developed  by
S.G.  Sergeev.  The  wavelength  calibration was based on the night sky lines 
and narrow emission lines of the  galaxy.
The spectra have been normalized to the  [OIII]$\lambda5007$ emission line 
flux.  

H$\alpha$ and additional H$\beta$ observations were performed on   August 7,
1998 with the 2.5 m INT at La  Palma. We used the 
Intermediate Dispersion Spectrograph (IDS) and the 235 camera in combination
with the R1200Y grating.  Two exposures of 1800 and 825 s,
respectively, included H$\beta$ and another one of 1800 s included 
H$\alpha$. The
seeing was $1''.1$ and the slit width $1''.5$. The spectral resolution was 1.8
\AA.  Standard reduction  procedures including flat-fielding, wavelength
calibration, spectral response,  and sky subtraction were performed with the
help of the IRAF software package. 

 The red-shift of III Zw 2  was taken to be  z=0.0898
(V\'eron-Cetty \& V\'eron 2000).

\section{Gaussian analysis}

\subsection{Line profile analysis} 

The first step in analyzing the emission lines was to define the
continuum. In the case of H$\beta$ the local continuum in four narrow zones
around  4800, 4900, 5520 and 5600  \AA\ was fit with a second order
polynomial.  For the H$\alpha$, Ly$\alpha$ and Mg II$\lambda2798$ lines
the
continuum was estimated using a straight line between two
wavelengths: around 6900 \AA\ and 7400 \AA\ for H$\alpha$;  1200\ \AA\ and
1450 \ \AA\ for Ly$\alpha$; and 2900 \AA\ and 3200\ \AA\ for the Mg II line.
We fit each line with a sum of Gaussian components using a $\chi^2$
minimalization routine to obtain the best fit parameters. We have also
assumed that the narrow emission lines can be represented by one or more 
Gaussian
components (see text below).  In the fitting procedure,  we look for the
minimal number of Gaussian components needed to fit the lines. It was 
found that
three broad Gaussian components provides a good fit to the profiles of
the 
Ly$\alpha$
and Mg II lines, while one additional narrow component was needed to fit 
the
Balmer lines. 

In Figs. 1--5 we can recognize clear evidence of substructure
in 
all the BELs. In the line profile with highest resolution (the H$\beta$
spectrum 
from the INT, see Fig. 1), the narrow emission line is blue-shifted with
respect  to the BEL which exhibits a gentle slope towards the red and a
steeper drop  towards the blue. Asymmetries are also present in
the H$\alpha$, Ly$\alpha$ and  MgII BELs (Figs. 2--4). The absence of a 
narrow 
emission-line component in the Ly$\alpha$ and  MgII lines  indicates that
the 
contribution of the Narrow Line Region emission in these lines is minor. 

\subsubsection{H$\beta$}

To study in more detail the presence of substructure  we performed a 
multi-Gaussian fit to the INT high resolution H$\beta$ line
profile. The low resolution and low S/N ratio of CrAO 
spectra could not be qualitatively decomposed into Gaussian components
and we use the spectra only for line variation investigation (see
Sec. 4). To 
limit the  number of free parameters in the fit we have set some
{\it a priori}  constraints (Popovi\'c et
al.
2001,2002). In the first place, the three narrow Gaussians representing the 
two 
[OIII]$\lambda\lambda 4959,5007$ lines and the narrow H$\beta$ component are 
fixed at the same red-shift with Gaussian widths proportional to
their  wavelengths. Full Width at Half Maximum (FWHM) is connected with width 
of the Gaussian profile (W) as $FWHM=2W\sqrt{\ln 2}$.  Second, we have linked  
the intensity ratio of  the two 
[OIII] lines according to  the line strengths, 1:3.03 (Wiese et al. 1966).  
Finally, we have
included  in the fit a red shelf  Fe II template   consisting of  nine Fe 
II  lines belonging to the multiplets 25,
36
and  42  (Korista 1992). We  took  the relative  strength of  these lines  
from Korista
(1992)  and supposed  that all the Fe II lines originated 
in the  same region,  that is, all of them have the same
red-shift and widths proportional to  their wavelengths. 

We obtained reasonably good fits by considering the above mentioned narrow
and  shelf components and three broad H$\beta$ components with red-shifts:
0.0856 (W=2200 km/s), 0.0898 (W=1900 km/s)  and 0.0950 (W=2800 km/s),
see Fig. 1.  The central broad component is relatively weak and,  in
principle, consistent fittings can be obtained with only two broad 
components  on the basis of minimal Gaussian component
assumption. Therefore we used F-test (Eadie et al. 1971) in order 
to compare the $\chi^2$ of two and three broad Gaussian models and 
find that the model with the central component leads to a
significant improvement in fit quality at the 99.8\% confidence
level. On the other hand, as we will
see,  the central component is present in the other hydrogen and MgII
lines in the
same way. 

We have also used the high resolution spectra from the INT to study in detail 
the [OIII] lines. To do this we have subtracted from the original spectrum
all  the $H\beta$ components (broad and narrow) and the Fe II shelf, obtaining
the  spectrum shown in Fig. 5. In this Figure we note that both [OIII]
lines  show very extended wings and can not properly be fitted by a Gaussian.
We also  notice that the wings are asymmetrical, being more gently  sloped
towards the  blue. We have  performed a multi-Gaussian fit to these lines
finding that  at least one relatively broad (W =410 km/s) and blue-shifted 
(0.0875) component  should be included to account for the extended wings (see
Fig. 5).  One can expect that other narrow lines in the optical
spectra 
have the same shape as the [OIII] ones and that the [OIII] line profile
can 
be used as template to fit these lines. But taking into account that the
other narrow
lines are very weak in comparison with the corresponding broad one (see
e.g. [NII] in the H$\alpha$ wavelength region) the asymmetry seen in
[OIII] will not affect significantly to the line profile and for the
purposes
of the paper the narrow
lines can be satisfactory fitted
with one single Gaussian. { On the other hand   we can not be sure that
all
the narrow lines are
emitted
under the same kinematical and
physical conditions. }

\subsubsection{H$\alpha$} 

To fit the H$\alpha$ line, we have assumed that the [NII]$\lambda\lambda 
6548,6583$ and the H$\alpha$ narrow component have the same red-shift and
Gaussian  widths proportional to their wavelengths. Taking into account that 
the
two  [NII] lines belong to the transition within the same multiplet we
assume an intensity ratio of 1:2.96 (see e.g. Wiese et al. 1966).
However, a simple inspection to
Fig. 2 shows that the  peak of the [NII]$\lambda$6548 line is higher
than the
peak of the [NII]$\lambda$6583 line. This may indicate the presence of a
blue-ward
asymmetric underlying component.
We have been only partially
successful
in reproducing the narrow  [NII]$\lambda$6548 line. This is not very
important at
this stage but we will need  to clean it from the line profile to fit the
disk
model in \S 5. For this reason  we have considered an additional arbitrary
narrow
component at $\lambda$6548 to totally remove this line.

In this case we also found good fits using in addition to these narrow 
components, three broad components for H$\alpha$ (see Fig. 2). As in the 
H$\beta$ case there is a central broad component (z=0.0898, W =1250 km/s) 
located  between two other broad components,  red-shifted (z=0.095, W = 2500
km/s) and  blue-shifted (z=0.0861, W =2330 km/s) ones. 
In the case of H$\alpha$, the broad central component contributes a larger
fraction of the integrated emission-line flux than does the
corresponding component in H$\beta$.

\subsubsection{Ly$\alpha$} 

The complex Ly$\alpha$ shape contains three absorption lines and two narrow
emission lines, Fig. 3. In order to clean the absorption lines as well as
the emission
satellite lines we assume that each one of them can be represented by a
Gaussian. The central absorption line is red-shifted  around 0.0888 and
probably arises from a Ly$\alpha$ auto-absorption. The feature at
 $\lambda\approx 1335$ \AA\ is  possibly an intrinsic NV absorption
(Crenshaw et
al. 1999). The very weak absorption component in the blue wing may be the
SiIII$\lambda$1207 line (see e.g. Laor et al. 1994,1995).
{ The two narrow emission lines in the blue wing are very close to the
SiII$\lambda$1195,1197 lines (Laor et al. 1994, 1995), but the observed
line profiles are too narrow, that would not be expected for these
permitted Si transitions. This feature can be identified as contribution
of geocoronal OI$\lambda$1302,1306 emission lines (see Eracleous
1998). The lines were fitted with two Gaussian and subtracted from the
Ly$\alpha$ blue  wing.}

In the red wing of the Ly$\alpha$ line there appear the
NV$\lambda\lambda$1239,1243 lines
(Wilkes \& Carswell 1982, Buson \& Ulrich 1990, Laor et al. 1994, 1995). In
order to subtract these lines we supposed that they come from the same
emission region, i.e. that they have the same w/$\lambda$ and intensity
ratio I(1238)/I(1242)=1.98 (Wiese et al. 1966).   

We needed three  broad components to perform the multi-Gaussian fitting of the
Ly$\alpha$ line (Fig. 3) with parameters: W $\approx$ 3150
km/s, z$\approx$  0.083; W $\approx$ 1380 km/s, z$\approx$
0.0898, and W $\approx$ 3250 km/s,  z$\approx$ 0.0943. The
width and red-shift of the Gaussians fitted to the NV lines were 2270 km/s and
z$\approx$ 0.0898, respectively. The estimated ratio of I(NV)/I(Ly$\alpha$)
is around 0.12, in a very good agreement with previous estimates (Laor et 
al. 1994, 1995).

\subsubsection{MgII$\lambda2798$} 

The contribution of 33 Fe II lines from multiplets 60, 61, 62 and 63  
to the red and blue wing of Mg II$\lambda2798$ has been taken into account. 
We assume that 
line intensities ratio  within a multiplet is proportional to the ratio of 
corresponding line strengths. The atomic data for the line strength we took 
from the NIST web site 
(http://physics.nist.gov/cgi-bin/AtData). We also assume that 
the Fe II emission  originates in the same region, i.e. the lines have the 
same width and shift.  The decomposition of MgII$\lambda2798$ is shown in 
Fig 4.  The scaled and broadened Fe II template
is indicated by the dashed  lines (bottom).
As in the case of the H$\beta$, H$\alpha$ and the Ly$\alpha$ lines, the MgII
line  can be decomposed into three broad Gaussian components (Fig. 4) with
parameters:   W $\approx$ 2700 km/s, z$\approx$ 0.0843;  W $\approx$
1450 km/s,  z$\approx$ 0.0896; and W $\approx$ 3600 km/s, z$\approx$
0.0980. The lines from the Fe II template have $W=2100$ km/s, $z=0.0898$.

\subsection{Discussion of the multi-Gaussian analysis} 

In Fig. 6 we  present the width  of the  different broad 
components 
versus their centroid velocities (relative to the systemic one). The
different 
components appear well separated in this diagram, showing the consistency of 
the multi-Gaussian decomposition.  By inspection of the diagram we can derive 
the following conclusions: 

(i) the best fit with Gaussian functions can be obtained only if we use  
three broad
Gaussians. 

(ii) the Gaussian decomposition  indicates the existence of a central
broad 
component of red-shift consistent with the systemic velocity. 

(iii) the presence of  red- and blue-shifted broad components in the case
of
all considered lines  suggests that  part of the  emission may originate in a
different region, possibly a disk.

If we assume that a disk (or a disk-like) region exists, we can roughly 
estimate the parameters of the disk using the results of Gaussian 
analysis 
and the relationship (see Popovi\'c et al. 2002)

$$\sin i\approx {\Delta z\sqrt{R_{\rm out}}},$$
where $i$ is the inclination of the disk, $R_{\rm out}$ is outer 
radius given in Schwarzschild radii
($R_{Sch}=2GM/c^2$). Taking into account that $\sin i\leq1$, we can 
estimate the 
maximal outer 
radius. From our
analysis we find that $\Delta z=z_R-z_B$ (where $z_R$ and $z_B$  are
the shift of the red
and blue Gaussians, respectively) is in the interval
from 0.0086 (for H$\alpha$) to 0.014 (for Mg II 2798 line). Then we can
estimate the maximal outer radius,
$R^{\rm max}_{\rm out}< 10^4R_{Sch}$. On the other hand, if we accept from
previous
investigations (e.g. Wandel et al. 1999, Kaspi et al. 2000,   Popovi\'c et
al. 2001, 2002) that the
outer radius of a BLR has typical dimensions of
 $\sim 1000 R_{Sch}$, we estimate that $i\sim 10^\circ-20^\circ$. This can 
be used as a starting point in analyzing the line shapes using a more  complex 
model of the BLR. The inclination obtained by us is significantly smaller than 
the one obtained  by Rokaki 
and Boisson (1999) 
for III Zw 2, but at the same time it is in agreement with a mean value of 
disk inclination of 21 Sy 1 galaxies given by these authors, as well as with 
their conclusion that "we tend to observe the Sy 1 galaxies at a more face-on 
inclination" (Rokaki and Boisson 1999).

\section{Variability} 

We  have used  the   long-term  H$\beta$  observations  to  discuss the 
variability of the BELs. With this aim we have removed from each observed 
spectrum  its  continuum  and used the emission in the narrow 
[OIII]$\lambda$5007 line to normalize  the spectra. The high-resolution
spectrum obtained with the INT, has been smoothed  to match spectral
resolution  of the CrAO spectra.  Five spectra of low
S/N were
rejected from the whole set. 

To find any variation in the H$\beta$ line  profile along the observed period
we constructed profiles of the mean, of the RMS, and of the RMS divided by the
square root of the mean (Figures 7abc). As one can see in Fig. 7a the mean
profile is single  peaked and asymmetric with traces of shoulders in the blue
and red wings which may represent substructure connected with emission from a
disk. The shoulders appear clearly enhanced in Figure 7b; a more prominent one
at $X \sim -0.008$ and the other two  maxima at $X \sim 0.006$ and 0.012,
respectively ($X=(\lambda-\lambda_0)/\lambda_0$).
 According to Figure 7c the
highest variations are found in the region including these features between
approximately $X=-0.01$ and $X=0.013$. 
 In principle, a  part of   
the variability in the red part of the H$\beta$ emission line might  be 
attributable to changes in the Fe II contribution, however we do not found
evidences of variability in the stronger Fe II component  beneath [OIII]
(see Figs. 7bc).
Leaving aside
the H$\beta$ profile, the strong dispersion in the residual He II line at $X
\sim -0.018$ is noticeable.


In order to study the line flux variation during the observed period,  we
have
presented in Figure 8a the integrated flux (H$\beta$ plus [OIII], Fe II and
He II lines) between 4750\AA\ and  5050\AA\ (rest wavelengths) for each one of
the individual spectra. The resulting integrated light curve decreases a $\sim$
50\% from 1972 to 1998.  To improve the S/N ratio we have considered seven sets
of spectra (defined according to the observational gaps, see Figure 8a)
averaging the spectra within each set. The averaged light curve (Fig. 8b)
reproduces the variation inferred from the individual spectra and  is in good
qualitative agreement with the slowly decreasing trend found by Salvi et al. in
the optical B-band (see Fig. 7 in Salvi et al. 2002). 

To study the flux variation of different parts of the line we have applied to
the seven averaged spectra the same Gaussian analysis made in \S 3.1.1. To
improve the fits we have, in first place, fitted the mean H$\beta$ profile
obtaining centroids for the three broad Gaussians: central ($\sim 0.0893$)
blue-shifted ($\sim 0.084$) and red-shifted ($\sim 0.0953$); similar to the
ones obtained  from the INT H$_\beta$ profile (\S 3.3.1.). In a second step we
performed the Gaussian fit of the seven averaged spectra fixing the Gaussian
centroids to these values. The results of this analysis can be seen in Figs. 8
and 9. In Fig. 8c we present the variation of the sum of the three Gaussians.
This light curve is practically the same as the one corresponding to the
integrated flux (Figure 8b). This confirms that the main contribution to the
flux variation in the considered wavelength range is the variation of the
H$\beta$ broad component. In Figs. 9abc, the light curves corresponding to each
one of the three Gaussian components are presented. The variation
in the blue and red Gaussians (with an amplitude greater than that of the
central Gaussian) tends to be correlated, once again supporting the
assumption of the existence of two
regions   contributing  differently  to the H$\beta$ broad
emission line profile.


\section{Two-component model analysis}

As mentioned in the introduction, the accretion disk model was taken
into 
consideration for III Zw 2 in previous investigations (Kaastra \&
Korte 1988, 
Shimura \& Takahara 1995, Rokaki \& Boisson 1999). The shape of the line 
profiles discussed in the present paper (two, red- and blue-shifted
broad components) and our preliminary analysis also support this idea. 
Besides, the multi-Gaussian fitting  also implies the presence of a 
central 
component with the systemic velocity. According to this result, in 
this
section we 
are going to fit the lines using a two-component model based on a disk 
and
a 
central Gaussian component which can be interpreted as a region surrounding the 
disk. 

For the disk we  use the Keplerian relativistic model of 
Chen 
\& Halpern (1989). The emissivity of the disk as a function of radius, $R$, is 
given by 
$\epsilon=\epsilon_0R^{-p}.$ 

Considering that the illumination is due to an extended source from the
center of the disk and that the radiation is isotropic, the flux from
the outer disk at different radii should vary as $r^{-3}$ (Eracleous \&
Halpern 1994), i.e. $p=3$ and that is value we
 adopt. We express the disk dimension 
in gravitational radii ($R_g=GM/c^2$, $G$ being the gravitational
constant, 
$M$  the mass of the central black hole, and $c$  the velocity of light). 
The  local broadening ($\sigma$) and shift ($z_{\rm Disk}$) within the disk 
have been taken into account (Chen \& Halpern 1989), i.e.
the $\delta$ function has been replaced by a Gaussian function (with the 
mentioned
parameters).

Before performing the fitting we have 'cleaned' the spectra by subtracting:
(i) the narrow lines from INT high-resolution H$\beta$ and H$\alpha$; (ii) the 
absorption
features, the narrow
emission in the blue wing, and the NV lines from Ly$\alpha$;  (iii) the 
narrow [OIII] lines and the Fe II template from H$\beta$ and (iv) the Fe II 
template from the blue and red wing of the Mg II line. It is striking  that
after this operation is done, the features associated with the disk are 
visible not only  in the asymmetrical wings of  Mg II but also in the red and
blue shoulders of the Ly$\alpha$ and  H$\alpha$ as well as in the triangular 
shape and the red shoulder of H$\beta$ (see Fig. 10). 
To compare the line profiles we present in Fig. 10a the intensities normalized 
to the peak ones {\it vs.} $X=(\lambda-\lambda_0)/\lambda_0$. As one 
can
see from Fig. 10a the  lines have  similar  shapes.

When a chi-square minimization including all the parameters at once is
attempted, it is found that the results are very dependent on the initial
values given to the parameters.  To overcome this problem we have, in the 
first
place, tried several values for the inclination using an averaged profile of 
all four  lines (Fig. 10b). We found  that  the best
che-by-eye fits can be obtained for values of  
$i\approx 
12^\circ$.  
 Accordingly we have fixed it
 to $i=12^\circ$ and performed a chi-square fitting of the other
parameters starting from suitable initial values. 
 The fit of the 
BELs wings strongly restricts the value of the inner
 radius and additional "local"
 broadening, i.e. random velocity of emission gas in the disk. This fact
(related to the emissivity
 dependence, $p\approx -3$) supports the validity of the determination of 
these
 parameter from the line profile fits.  We note here
that changing the inner (outer) radius of the disk and parameter $p$ 
we can
obtain a satisfactory fit with the inclination $12^\circ\pm 5^\circ$.

The results of the fit are presented in Fig. 10bcdef, and the 
disk and  Gaussian
parameters  in Table 1.   
 This  Table enables us to point out the following
results:  (i) There is a very good   consistency  among the parameters (z 
and
W) of
the broad  components representing the region surrounding the disk. 
Their
red-shifts exhibit  a very small difference  with 
respect to the red-shift of  the [OIII] narrow lines. (ii) There is also a
good consistency in the red-shifts   for the disk corresponding to 
Ly$\alpha$,
H$\alpha$, H$\beta$, and MgII. The average $z$ for these four disk lines
appears to be slightly blue-shifted (by about 600 km/s) with respect to the
systemic one.   
(iii) The inner radius of the
Ly$\alpha$ 
emitting disk is clearly  smaller than the others. H$\alpha$ and H$\beta$
exhibit a very good coincidence of the inner radii but the inner edge of 
the
MgII emission ring seems to  be closer to the  disk center although this point
should be viewed with caution. (iv) The disk emission component contributes 
more to the total flux than the low-velocity component of the BLR.

Taking into account the estimated mass of the central object in III Zw 2 
(M$\sim 2\cdot 10^{8}\ M_\odot$)  given by Vestergaard (2002) we can
obtain the  dimensions of the radiating  disk: $R_{\rm inn}\sim
5\cdot
10^{13}$ m,  $R_{\rm out}\approx 3\cdot 10^{14}$ m. This last value is in 
agreement
with the  estimation  given by Kaastra \& Korte ($\sim 2\cdot10^{14}$ m). The
size of the whole BLR  (disk + surrounding region) cannot be inferred from this
analysis but it might be  considerably larger (Collin \& Hur\'e 2001). On
the other hand, the BLR surrounding the disk may  originate from an
accretion disk wind, that may be created due to several disturbances
capable of producing shocks (e.g. Bondi-Hoyl flow, stellar wind-wind
collision, and turbulences, see e.g. Fromerth \& Melia 2001).
 Also, a Keplerian disk with disk wind can produce single
peaked broad emission lines (Murray \& Chiang 1997).
 Recently, 
 Fromerth \& Melia (2001)   described a scenario of the
formation of BLR in the accretion disk shocks, that can create a
surrounding BLR. 

We should mentioned here that other geometries can  contribute to the
substructure seen in III Zw 2 line shapes.
Besides  emission of the  disk (or disk-like region) or emission from
spiral shock waves within a disc (Chen et al., 1989, Chen \& Halpern, 1989),
the following geometries may cause substructures in line profiles:
i) emission from the oppositely-directed
sides of a bipolar outflow (Zheng et al., 1990, Zheng et al., 1991);
ii) emission from a spherical system of clouds in
randomly inclined Keplerian orbits, illuminated anisotropically from the
center (Goad \& Wanders, 1996); and
iii) emission of the binary black hole system (Gaskell, 1983, 1996). But, in
any case  the two-component model should be taken into account, considering
a low-velocity BLR and one additional emitting region.

\section{Conclusions} 

We have analyzed UV spectra and a collection of optical spectra of III Zw
2 procured in over 20 years. The flux of H$\beta$ spectra shows 
variability in
the  wings, as well as in the line core. The variation of the blue and red wing
 fluxes tends to correlate during the considered period. It indicates that
line wings
 originate in the same region, while the line core arises from another
emission line region (low-velocity BLR).
  We have also discussed the possible 
contribution of a Keplerian disk of emitters to the BELs, finding the 
following results: 

1 - The shape of the BELs (especially after removing the narrow and
absorption 
lines) indicates  a multicomponent origin, and certain 
features --like 
 the shoulders in Ly$\alpha$, H$\alpha$ and H$\beta$ and the slight
profile asymmetries -- that 
can be associated with a disk. 

2 - The same two-component model (Keplerian relativistic disk +
a surrounding emission region) can consistently fit the 4 BELs considered
here 
(Ly$\alpha$, MgII, H$\beta$, and H$\alpha$). This   supports  the 
standard model hypothesis in the sense that a part of the broad line
emission arises from a Keplerian disk. 

3 - From the fitted disk parameters and the mass of the central object
(Vestergaard 2002) we can estimate that the Ly$\alpha$ disk has inner
and outer  radii of around 0.0018 and 0.01 pc, respectively. 
However, the inner radius is greater for the
 Mg II ($\sim 0.0027$ pc) and for the H$\alpha$ and H$\beta$ lines ($\sim
0.0038$
pc). This indicates a radial stratification in the disk.
 The
relatively broad component present in the  blue wings of the narrow [OIII]
lines is another indication of stratification  and perhaps could indicate a
connection between the outer BLR and the NLR.

\section{Acknowledgments}

 This work was supported by the Ministry of
Science, 
Technologies and Development of Serbia through the project P1196 
``Astrophysical
Spectroscopy of 
Extragalactic Objects'' and  the project P6/88 "Relativistic and
Theoretical Astrophysics" supported by the IAC.
 L. \v C. P. is supported by Alexander von Humboldt Foundation through the
program for foreign scholars.
 L. \v C. P. and E. B.  thanks the
Institute for Astrophysics 
Canarias for the hospitality during his stay at the Institute. 
We would like to thank to the anonymous referee for the
very useful comments.

\clearpage

\begin{table*} \caption{The parameters of disk: z$_{\rm disk}$ is the
shift and $\sigma$ is the  Gaussian broadening term
from disk indicating the random velocity in disk, R$_{\rm inn}$ are the
inner
radii, R$_{\rm out}$ are the outer radii. The z$_{\rm G}$ and W$_{\rm
G}$ represent the parameters of the Gaussian component.  $<AV>$
is an averaged profile (see Fig. 10b). $F_D/F_G$ represents the ratio of the 
relative disk and Gaussian fluxes. }

\vbox{\tabskip=0pt   \offinterlineskip  \def\podvuci{\noalign{\hrule}}
\def\razmak{\noalign{\vskip.1cm}}     \halign    to    \hsize{\strut#&
\vrule#\tabskip=0pt  plus   10pt  minus5pt  &
 \hfil#\hfil&  \vrule#&
 \hfil#\hfil&  \vrule#&
\hfil#\hfil&  \vrule#&
\hfil#\hfil&  \vrule#&
 \hfil#\hfil&  \vrule#&
\hfil#\hfil&  \vrule#&
 \hfil#\hfil&  \vrule#&  \hfil#& \vrule #
  \tabskip=0pt\cr\podvuci && Line && z$_{\rm disk}$  &&$\sigma$
(km/s) &&
R$_{\rm inn}\ (R_g)$   &&  R$_{\rm out}\ (R_g)$  && z$_{\rm G}$ &&
W$_{\rm G}$ (km/s)&& $F_D/F_G$
&\cr\podvuci\razmak\podvuci

&& Ly$\alpha$&& -800 && 850 &&  200 && 900 && -20 && 1280&& 1.11  &\cr
&& Mg II$\lambda$2789 && -350 && 920 && 300 && 1000 && -30  && 1100&& 1.86 
&\cr
&&H$\beta$  && -600 && 920 && 400 && 1300 && -130 && 1100&& 3.14 &\cr
&& H$\alpha$ && -600 && 850 && 450 && 1300  && -120 && 1170&& 1.52 &\cr
&& $<AV>$ && -600 && 890 && 400 && 1200 && -120  && 1170&& 1.72 &\cr
\podvuci}}
\end{table*}

\clearpage

\begin{figure}
\resizebox{8.2cm}{!}{\includegraphics{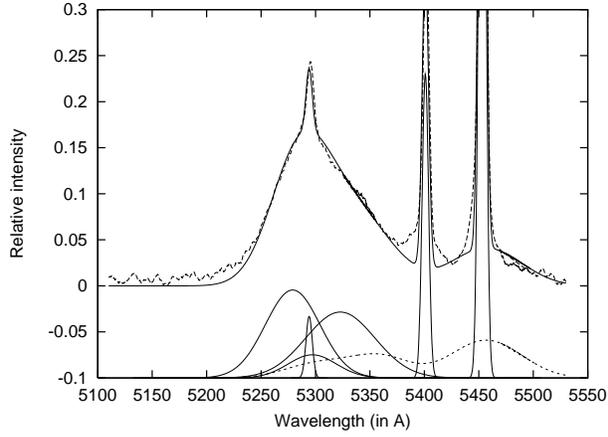}}
\caption{ Decomposition of the
H$\beta$ line observed with INT.  The dashed line represents observations
and solid line shows the profile
 obtained by Gaussian decomposition.  The Gaussian components are 
presented
at bottom.
 The dashed complex line, at bottom, represents the contribution of the Fe 
II lines } \end{figure}

\begin{figure}
\resizebox{8.2cm}{!}{\includegraphics{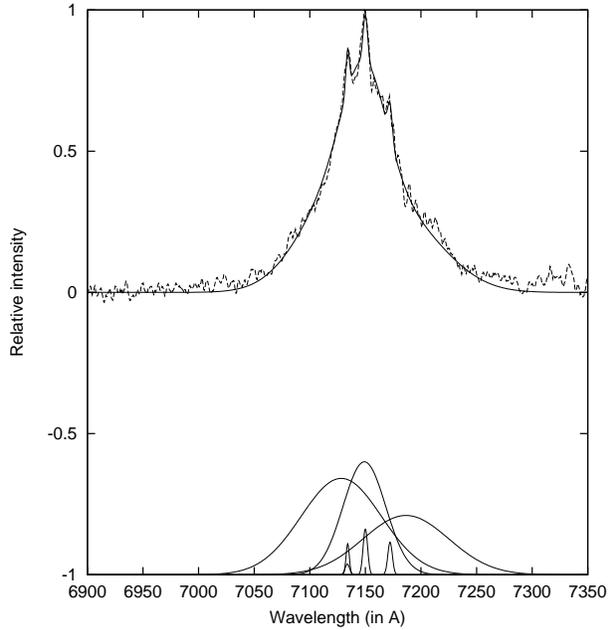}}
\caption{Same as in Fig. 1,
but for the H$\alpha$ line. Besides the narrow H$\alpha$ central Gaussian, 
two weak [NII] lines have been detected: at $\lambda\lambda$7136 \AA\ and 
$\lambda\lambda$7174 \AA\ (see  text for details)
} \end{figure}

\begin{figure}
\resizebox{8.2cm}{!}{\includegraphics{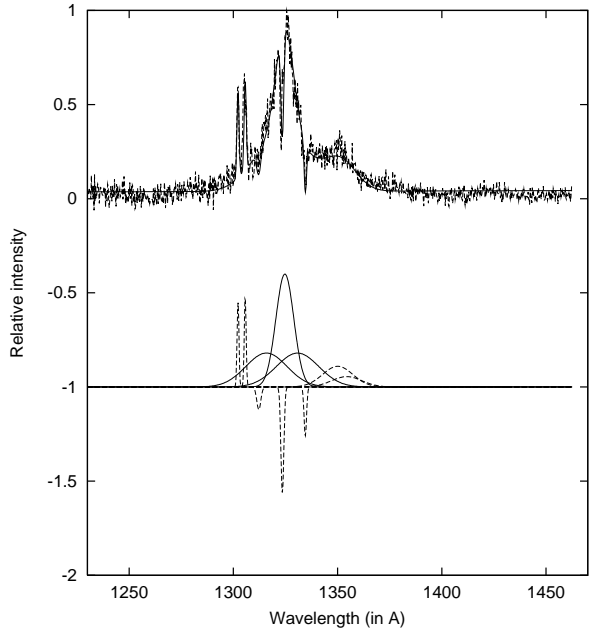}}
 \caption{Same as in Fig. 1, but for the
averaged shape of the Ly$\alpha$ line. The dashed lines, at the bottom,
represent the contribution of the narrow emission and absorption lines
(see
 text for details).}\end{figure}

\begin{figure}
\resizebox{8.2cm}{!}{\includegraphics{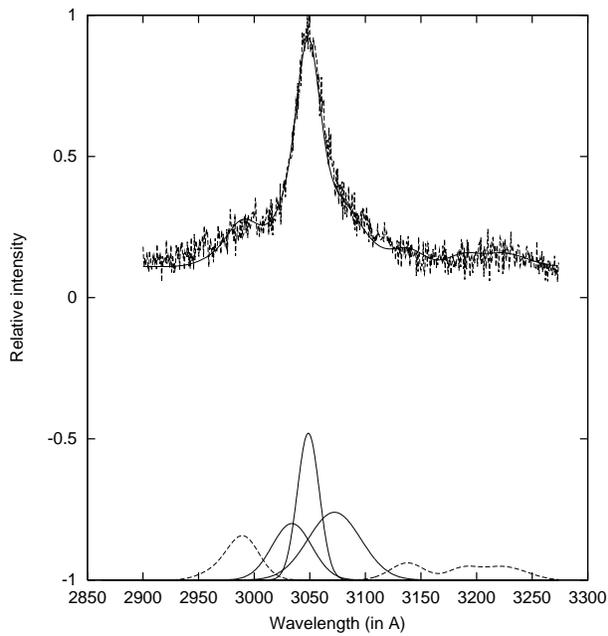}}
\caption{Same as in Fig. 1, but for
the Mg
II[2798] line. The dashed complex lines, at the bottom, represent the 
contribution 
of the Fe II template.} \end{figure}

\begin{figure}
\resizebox{8.2cm}{!}{\includegraphics{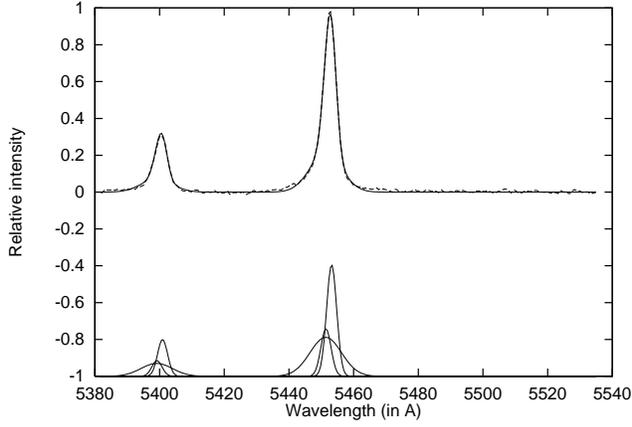}}
\caption{[OIII] Gaussian fitting and decomposition}
\end{figure}

\begin{figure}
\resizebox{8.2cm}{!}{\includegraphics{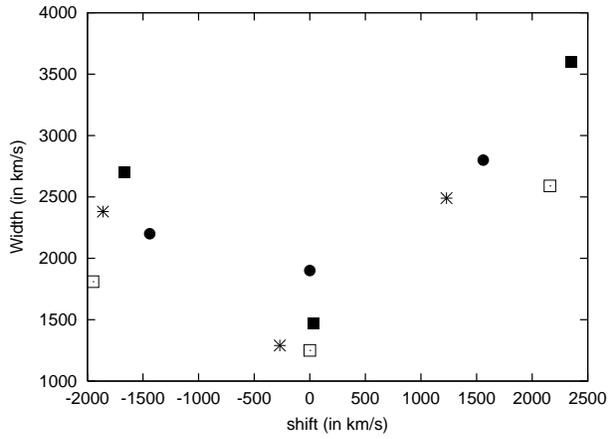}}
 \caption{The width, W, of Gaussians as a
function of the shift. The used notation is: full squares for
Ly$\alpha$,
 full circles for H$\beta$
observed with INT,
stars for H$\alpha$ and open squares for Mg II line}
\end{figure}

\begin{figure}
\resizebox{8.2cm}{!}{\includegraphics{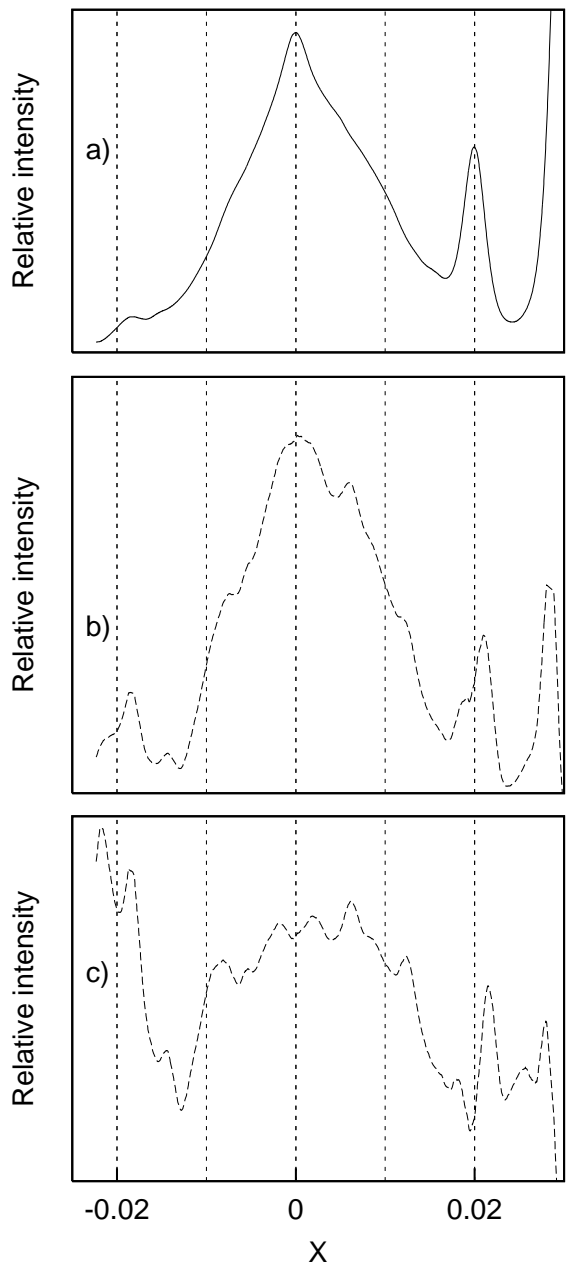}}
\caption{The mean H$\beta$ line profile obtained from 34 (33 from CrAO and
1 from INT) spectra (a); the
corresponding RMS profile (b) and the RMS divided by the square root of
the mean profile. 
The 
intensity scale for H$\beta$ line is from 0.0 to 0.5; for H$\beta$ RMS
is 
from 0.0 to 0.06 and for RMS divided by the square root of the mean
profile is from 0.0 to 0.1 in units of [OIII]5007 line flux. The value
$X$ is $(\lambda-\lambda_0)/\lambda_0$.}
\end{figure}

\begin{figure}

\resizebox{8.2cm}{!}{\includegraphics{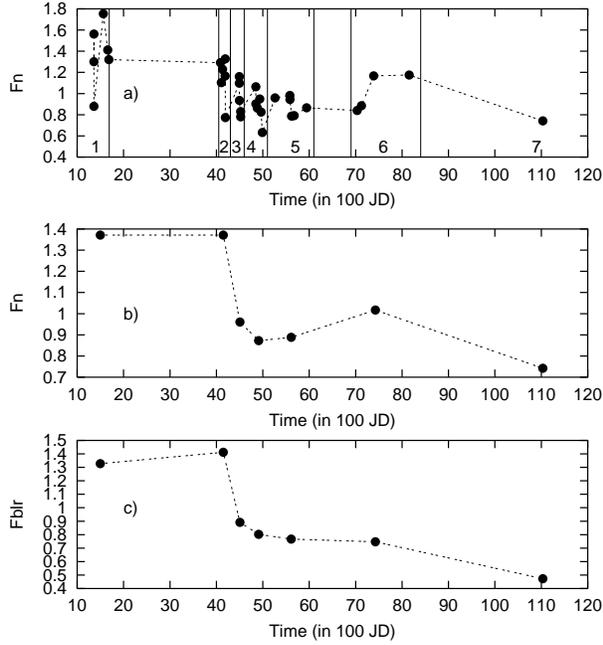}}
\caption{The  variation of: a) the flux ($F_{n}$) of spectra in H$_\beta$
line
region
(H$\beta$ plus [OIII]
and Fe II lines) normalized to the flux  of mean spectra in
H$_\beta$ line
region;  b) the
same variation of averaged spectra of the seven considered
groups; c) the total broad line flux ($F_{blr}$ - summ of fluxes of red,
blue
and central
componets obtained from
Gaussian
analysis) of seven considered groups normalized to the corresponding
broad component
flux of mean H$\beta$
line. The time is given
in 100 JD, starting from the epoch 2440000 (May 23, 1968).}
\end{figure}

\begin{figure}
\resizebox{8.2cm}{!}{\includegraphics{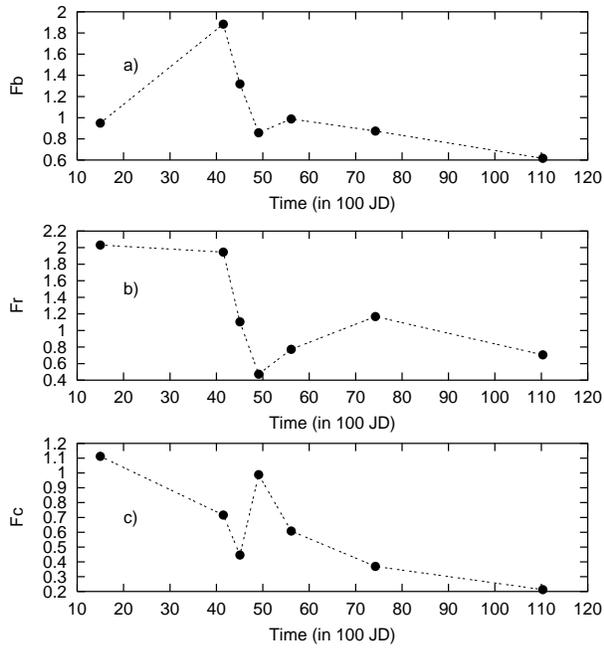}}
\caption{The  flux variations of: (a) the blue - $F_b$; (b) red $F_r$ and
(c) central - $F_c$ H$\beta$ 
component
normalized to the  corresponding component from the mean H$\beta$
line. The time is
given in 100 JD, starting from the epoch 2440000.}
\end{figure}

\begin{figure}
\resizebox{14cm}{!}{\includegraphics{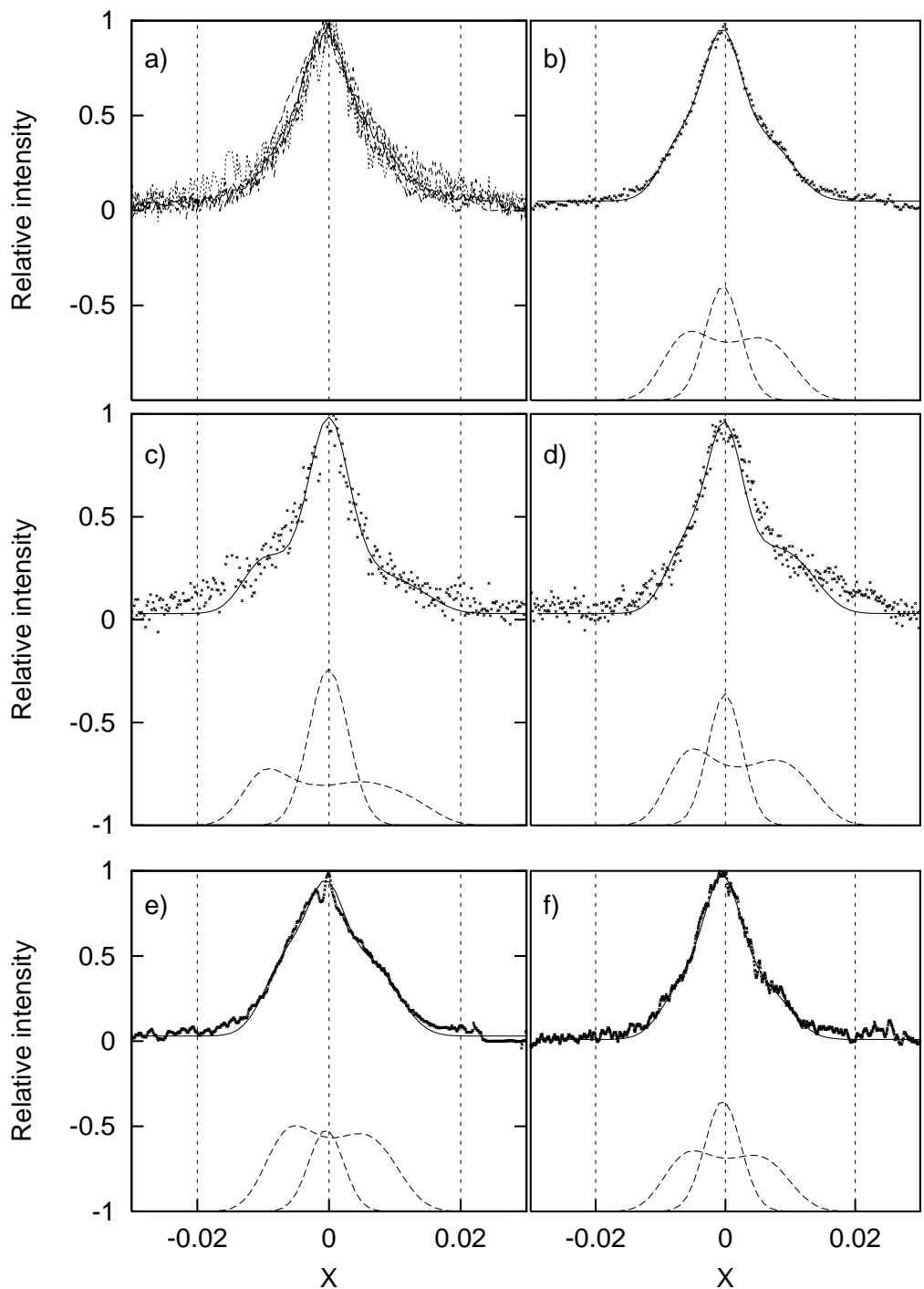}}
 \caption{Observed lines 
 of III Zw 2 (dots),  fitted with the disk model
(double-peaked) and one Gaussian; a) the comparison of all line profiles 
(dashed lines) with an averaged one (solid line); 
b) fit of the averaged line profile. Panels c,d,e,f represent fit of 
Ly$\alpha$, Mg II, H$\beta$ and H$\alpha$ lines, respectively. The value
$X$ is $(\lambda-\lambda_0)/\lambda_0$.  } 
\end{figure}

\end{document}